# Improving quantum-transition temperatures in BaFe$_2$As$_2$-based crystals by removing local-lattice strain & electronic-structure disorder

Li Li,[1,*] Qiang Zheng,[1,*] Qiang Zou,[2,*] Shivani Rajput,[1,2] Anota O. Ijaduola,[3] Zhiming Wu,[2,4]

Xiaoping P. Wang,[5] Huibo B. Cao,[6] Miafang Chi,[2] Zheng Gai,[2] David Parker,[1] Athena S. Sefat[1,^]

[1] Materials Science and Technology Division, Oak Ridge National Laboratory, Oak Ridge, TN 37831, USA
[2] Center for Nanophase Materials Sciences, Oak Ridge National Laboratory, Oak Ridge, TN 37831, USA
[3] Department of Physics, University of North Georgia, Dahlonega, GA 30597, USA
[4] Fujian Provincial Key Laboratory of Semiconductors and Applications, Collaborative Innovation Center for Optoelectronic Semiconductors and Efficient Devices, Department of Physics, Xiamen University, Xiamen 361005, P. R. China
[5] Chemical and Engineering Materials Division, Oak Ridge National Laboratory, Oak Ridge, TN 37831, USA
[6] Quantum Condensed Matter Division, Oak Ridge National Laboratory, Oak Ridge, TN 37831, USA

[*] *Authors contributed equally to this manuscript*
[^] *Corresponding author: sefata@ornl.gov*

Quantum materials such as antiferromagnets or superconductors are complex in that chemical, electronic, and spin phenomena at atomic scales can manifest in their collective properties. Although there are some clues for designing such materials, they remain mainly unpredictable. In this work, we find that enhancement of transition temperatures in BaFe$_2$As$_2$-based crystals are caused by removing local-lattice strain and electronic-structure disorder by thermal annealing. While annealing improves Néel-ordering temperature in BaFe$_2$As$_2$ crystal ($T_N$=132 K to 136 K) by removing in-plane electronic defects and overall reduction of $a$-lattice parameter, it increases superconducting-ordering temperature in optimally cobalt-doped BaFe$_2$As$_2$ crystal ($T_c$=23 to 25 K) by precipitating-out the cobalt dopants and giving larger overall $a$-lattice parameter. Although annealing promotes local chemical and electronic uniformity resulting in higher $T_N$ in the parent, it results in nanoscale phase separation in the superconductor resulting in lower disparity and strong superconducting band gaps in the dominant crystalline regions, which lead to both higher overall $T_c$ and critical-current-density, $J_c$.

After much research on cuprates and iron-based superconductors, the causes of unconventional superconductivity, and at a particular $T_c$, remain elusive. Iron-based superconductors can be created by chemical substitutions and charge doping of antiferromagnetic 'parent' materials that alter the nuclear and electronic structures, Fermi surfaces, carrier concentrations, and strength of spin fluctuations [1-8]. As is well-known, $A$Fe$_2$As$_2$ ($A$= Ca, Sr, Ba) parents (known as '122s') have coupled antiferromagnetic striped order (Fe spins parallel along $b$-, antiparallel along $a$- and $c$-axes) and orthorhombic transitions below $T_N$ =$T_s$ [9]. However, there are variations of $T_N$ in the literature within each family. For example, a BaFe$_2$As$_2$ crystal is reported to have $T_N$ = 134 K [10,11] that rises to $T_N$ = 140 K after thermal annealing (at 700°C) [12,13], while polycrystalline samples have $T_N$ = 140 K [14]. Similarly, $T_N$ values in SrFe$_2$As$_2$ can vary by 6 to 8 K [13-15]. Moreover, CaFe$_2$As$_2$ crystals can be produced to have different $T_N$ and even a non-magnetic ground state [13,16]. We found that the thermal-annealing temperature matters and it can lead to homogenous crystalline lattices giving the highest coupled $T_N$ =$T_s$ [13]. As there are several parents and superconducting doping types, the cause of superconductivity for a particular dopant and at a particular superconducting temperature is unpredictable [1]. Even looking within the family of BaFe$_2$As$_2$-based crystals and a particular doping level (x), $T_c$ variations can be observed. For example, the specific heat results on Ba(Fe$_{1-x}$Co$_x$)$_2$As$_2$ with x=0.045, 0.08, and 0.105 have shown that annealing (800 °C, 2 weeks) increases $T_c$ between 2 to 5 K, decreases the residual linear term in specific heat ($\gamma_o$) in the superconducting state by as much as half while suppressing the Schottky-like contribution below 1 K [17,18]. Moreoever, the magnetic susceptibility on Ba(Fe$_{1-x}$Co$_x$)$_2$As$_2$ crystals with 0.04≤x≤0.14 showed an increase of $T_c$ values by ~ 1 to 3 K, with no significant change in superconducting Meissner or shielding fractions [19]. In addition, annealing (800 °C, 1 week) of Ba$_{0.5}$Sr$_{0.5}$(Fe$_{1-x}$Co$_x$)$_2$As$_2$ with x=0.14 showed a $T_c$ increase of 5 K in bulk properties, and a decrease in heat capacity $\gamma_o$ of more than half [20]. In all of these examples, it is assumed that annealed crystals have improved crystallinity due to the release of residual strain, hence improve $T_N$ or $T_c$. This paper offers an insight on the complexity of Ba(Fe$_{1-x}$Co$_x$)$_2$As$_2$ quantum materials with intermingled effects of disorder, charge doping, and and electronic and crystal structure effects demonstrated by comparing results across multi-length scales using bulk techniques (diffraction, transport, magnetization), local probes (spectroscopy, microscopy) and theoretical input. Our results suggest that annealing improves electronic uniformity and antiferromagnetic order, while it promotes clustering of cobalt dopants at nanoscales to form more pinning sites and improved $J_c$, yet uniform and stronger areas of superconducting gaps that give improved $T_c$.

We assume that the overall strength of antiferromagnetism or superconductivity below a particular transition temperature in a crystal is produced by a complex combination of many local-scale details such as chemical and lattice structures, defects, and and electronic structure variations [1-7,21-23]. In this manuscript, we report on two sets of crystals, each with same average composition of x that give slightly different $T_N$ or $T_c$ values. Previously, we found praseodymium clustering to prevent bulk superconductivity in Pr-doped CaFe$_2$As$_2$ [24], and here we explore each pair of "as-grown" versus "annealed" antiferromagnetic BaFe$_2$As$_2$ or superconducting optimally cobalt-doped BaFe$_2$As, in order to understand reasons for their improved transition temperatures with annealing. We analyze the bulk properties in crystals, and also their nanoscale variations in the atomic-resolved and real-space lattice and electronic structures that get averaged by them. We report that higher transition temperatures in annealed crystals are due to the higher overall electronic and chemical uniformity, as expected. However, surprisingly, some cobalt dopants precipate out of crystalline matrix that yields larger electronically connecting regions with stronger superconductivity, and better pinning.

Although there is no distinct change in the average compositions of each as-grown versus annealed Ba(Fe$_{1-x}$Co$_x$)$_2$As$_2$ crystal, the average structures change only in-plane as will be shown here in $a$-lattice parameter variations, and smearing effects along the $ac$ plane. **Figure 1a** (top) shows typical size and quality of Ba(Fe$_{1-x}$Co$_x$)$_2$As$_2$ crystals that were used for our annealing studies, and the room-temperature tetragonal crystal structure that is made of covalently-bonded layers of (Fe/Co)As in the $ab$-plane separated by Ba ions along the $c$-axis. The average chemical composition of each crystal was measured with a Hitachi S3400 scanning electron microscope operating at 20 kV, and use of energy-dispersive x-ray spectroscopy (EDS). For the line analysis on a crystal, the instrument used was the Hitachi S3400 Scanning electron Microscope operating at 20kV; the beam current was set to provide approximately 1500 counts/second using a 10 mm sq EDAX detector set for a processing time of 54 microsecond; data were reduced using EDAX's standardless analysis program. **Figure 1a** (bottom) shows the variation in cobalt composition across a crystal with an average composition of 2.4% cobalt; this result shows that cobalt amount is non-uniform on the micrometer scale changing by ~0.1%. For all the crystals studied here, the x chemical composition is reported after averaging the results of energy on 3 random spots (~90 μm diameter each). For structural changes, powder X-ray diffraction data on numerous Ba(Fe$_{1-x}$Co$_x$)$_2$As$_2$ were collected on an X'Pert PRO MPD diffractometer (Cu $K_{\alpha 1}$ radiation, λ=1.540598 Å); the lattice parameters were refined by least-squares fitting within the program package *WinCSD* [25]. **Figure 1b** shows the refined lattice paramters versus x, from X-ray diffraction of powdered crystals. As expected for as-grown Ba(Fe$_{1-x}$Co$_x$)$_2$As$_2$ crystals [10,26], $a$-lattice constant remains mainly invariant while $c$-axis shrinks with increasing x due to smaller Co ions substituting for Fe. However, upon annealing, the $c$-lattice parameter does not change while the $a$-lattice parameter changes slightly, decreasing for the parent while increasing for optimally-doped x=0.063 crystal. This negligent $c$- parameter change of less than 0.01 Å was reported for x=0 annealed crystals [13]. To supplement these structural results, single crystal X-ray diffraction data on the set of as-grown and annealed crystals, with either x=0.063 or x=0.146, were collected on a Rigaku Pilatus 2000K diffractometer (Mo $K_\alpha$, λ=0.71073 Å). Crystals were mounted on MiTegen loops with a superglue for data collection at room temperature, with approximate sizes of ~0.1×0.1×0.02 mm$^3$. Data processing and reduction were carried out using the CrystalClear [27] software package. Crystal structure of the parent compound BaFe$_2$As$_2$ with a fixed site occupancy ratio for Fe and Co atoms at the 4*d* site was used as the starting model in structural refinement; structures were refined to convergence using SHELX-2014 [28]. **Figure 2a** shows evidence of peak broadening and lattice distortions for two sets of annealed crystals from axial photographs of single-crystal X-ray diffraction. We show the simulation for an overlay of a section of [101] layer to produce this smearing effects of lattice due to these results: for x=0.063, a small misalignment along [010] of ~ 1.0 degrees can give the evident peak broadening; for x=0.146 a twin law rotation along [010] by 2.4 degrees can produce peak splits. This result is also reflected in atom displacement along the crystallographic *c* direction, demonstrated in **Figure 2b**, found by refining the full dataset for x=0.063. The amplitude of this distortion causes the increase of arsenic height by ~ 0.0026 Å for annealed x=0.063, although the averaged *c*-lattice parameter stays the same. The smearing effect along the *ac* plane may be due to clustering of cobalt atoms in small regions as is evident from local microscopy results (explain below).

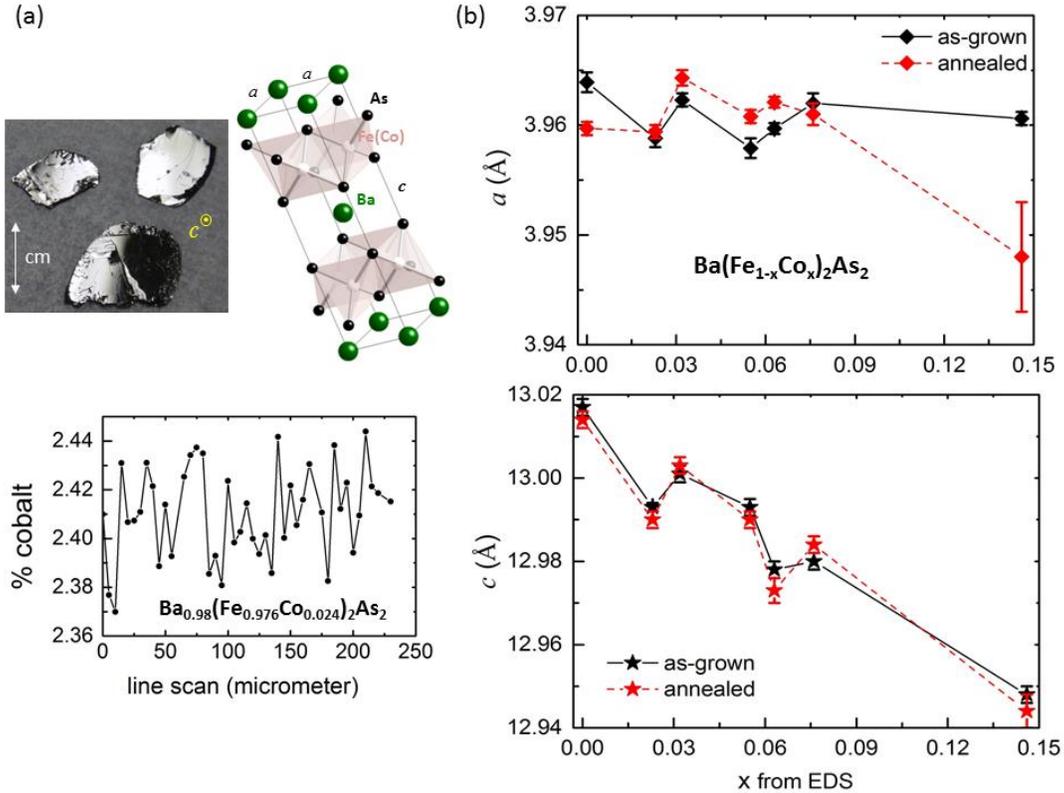

**Figure 1: Small variations of cobalt concentrations within a crystal, and changes in *a*-lattice parameters with annealing of Ba(Fe$_{1-x}$Co$_x$)$_2$As$_2$ crystals.** (a) (Top) Typical size and quality of as-grown Ba(Fe$_{1-x}$Co$_x$)$_2$As$_2$ crystals, and room-temperature tetragonal crystal structure of Ba(Fe$_{1-x}$Co$_x$)$_2$As, with unit cell shown in solid black line; (Bottom) A line scan across a crystal with average composition of Ba$_{0.98}$(Fe$_{0.976}$Co$_{0.024}$)$_2$As$_2$, assuming full occupancy of layers. (b) The refined *a*- and *c*-lattice parameters at room temperature, from powdered crystals with doping levels (x). The level of x does not change upon annealing, within error.

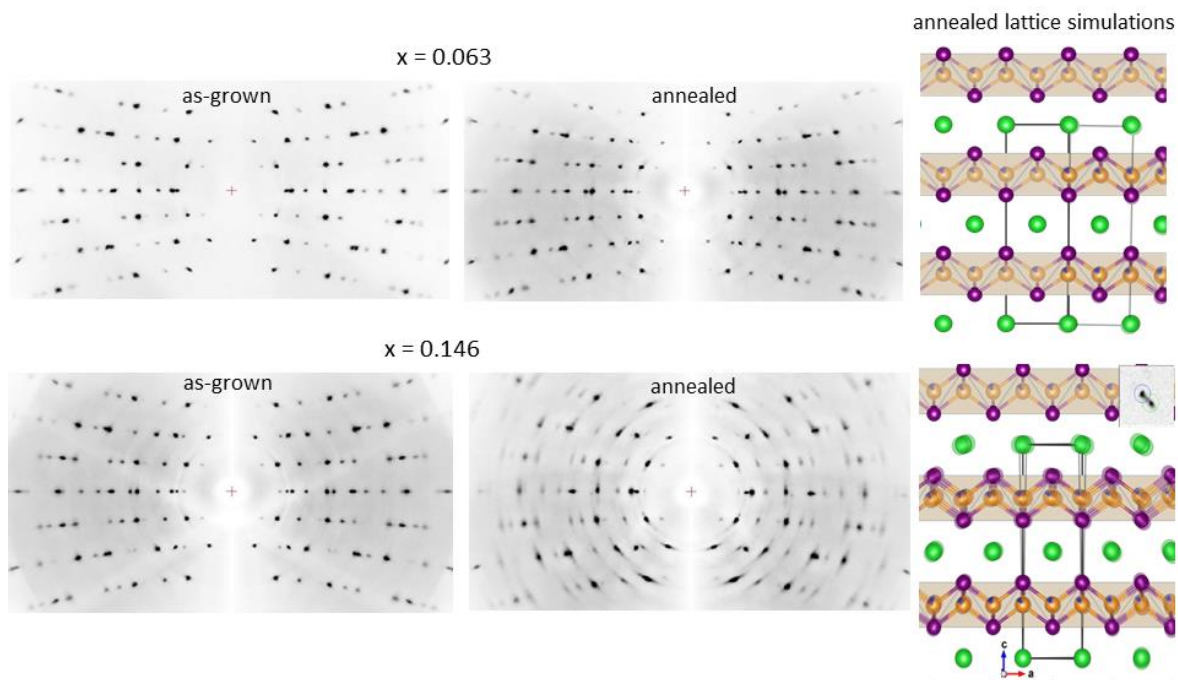

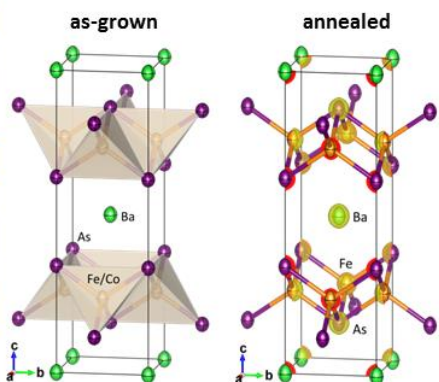

**Figure 2: Smearing effects in annealed crystals demonstrated by single crystal X-ray diffraction on Ba(Fe$_{1-x}$Co$_x$)$_2$As$_2$ crystals.** (a) (Left) [110] axial photographs computed from omega-scan images on as-grown compared to annealed crystals, for x=0.063 and 0.146. (Right) Overlay a section of [101] layers to simulate the effects of lattice misalignment for the annealed crystals: Top is demonstrating misalignment along [010] by ~1.0 degrees for x=0.063; bottom is showing twin law rotation along [010] by ±2.4 degrees, with upper-right giving the peak split, which may result for x =0.146. (b) Table shows the refined atomic displacement parameters[1] (Å$^2$) for as-grown and annealed crystals of Ba(Fe$_{1-x}$Co$_x$)$_2$As$_2$ with x=0.063. A site occupation ratio of 0.937:0.063 obtained from elemental analysis for Fe/Co atoms at 4$d$ site was used in structural refinement. Polyhedral drawing of the crystal structure of these crystals at the 90% ellipsoid level, is shown on the right. For annealed crystal, there is an overlay of electron density (e Å$^{-3}$) with iso-surface level shown at 5% of the maximum.

For the annealed BaFe$_2$As$_2$ crystal and as evident below, the coupled antiferromagnetic ($T_N$) and structural ($T_s$) transition temperatures improve while the crystal becomes more homogeneous. For measurements of bulk properties and inferring the transition temperatures, temperature-dependent magnetic susceptibility ($\chi$) data were collected on Ba(Fe$_{1-x}$Co$_x$)$_2$As$_2$ using a Quantum Design Magnetic Property Measurement System (MPMS), in zero-field-cooled or field-cooled modes, with field perpendicular to *ab*-plane at 10 Oe or 1 Tesla. Temperature-dependent electrical resistivity ($\rho$) or heat capacity ($C$) data were collected using a Physical Property Measurement System (PPMS). Hall coefficient ($R_H$) was calculated from the antisymmetric part of the transverse voltage perpendicular to the applied current under magnetic field ±6 T reversal at fixed temperature. **Figure 3a** represents these results on BaFe$_2$As$_2$. In the bulk properties, the shift of anomalies due to antiferromagnetic transition from $T_N$ = 132 K to 136 K is similar to those reported [13]. There is no change in the heat capacity Sommerfeld coefficient ($\gamma_o \approx 6$ mJ/K$^2$mol) upon anneal. The $R_H$ results on as-grown crystal is similar to those reported [29,30], although its magnitude increases for annealed crystal probably due to higher electron mobility. For finding the structural and magnetic ordering, single-crystal neutron diffraction was performed on ~ 0.02 gram pieces of the parent, measured at the four-circle diffractometer HB-3A at the High Flux Isotope Reactor at ORNL; neutron wavelength of 1.546 Å was used from a Si-220 monochromator [31]. In **Figure 3b**, the neutron diffraction results for the changes in nuclear and magnetic structures are shown. Comparing the tetragonal, T, (220)$_T$ Bragg peak for BaFe$_2$As$_2$ at 4 K and 200 K, there was not sufficient resolution to observe the peak splitting similar to others reported [32,33]. However, the intensity change with temperature due to peak broadening or extinction effect indicates the tetragonal-orthorhombic structural transition ($T_s$), as seen by the temperature dependence of peak intensity. The stronger intensity change in annealed crystal indicates the larger tetragonal-orthorhombic lattice distortion compared to as-grown. The peak intensities of the magnetic Bragg reflection (½½5)$_T$ versus temperature are also plotted here. We confirm that structural and magnetic transitions start at 132 K while for annealed crystal they occur at ~136 K, similar to transitions inferred from bulk properties. Our neutron diffraction results are also consistent with the combined result of magnetic susceptibility and X-ray diffraction measurements that showed that increase in structural and magnetic phase transitions are coincident [34].

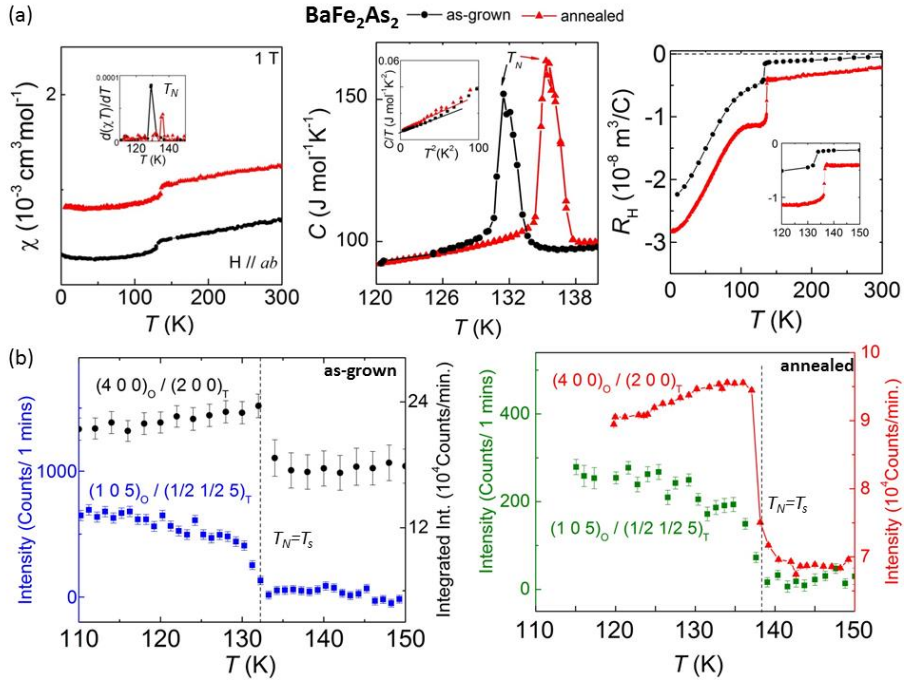

**Figure 3: Increase in the coupled antiferromagnetic ($T_N$) and structural ($T_s$) temperatures upon annealing BaFe$_2$As$_2$ crystal.** For BaFe$_2$As$_2$ crystals of as-grown (in black) and annealed (in red), temperature dependence of anomalies in (a) magnetic susceptibility ($\chi$), heat capacity ($C$) and Hall coefficient ($R_H$), and (b) structural and magnetic Bragg neutron reflections. In (b), temperature-dependence of the peak intensity change of the tetragonal (½½5)$_T$ magnetic reflection and (200)$_T$ structural Bragg peaks are shown.

The annealing of BaFe$_2$As$_2$ results in a more electronically homogeneous crystal, which is most probably due to lattice strain relief, confirmed here from scanning-tunneling microscopy/spectroscopy (STM/S) statistical analyses here on local areas. STM/S experiments were carried out with mechanically cut Pt-Ir tip in ultra-high vacuum variable temperature-STM chamber. The x=0 crystals of as-grown and annealed were mounted on a moly plate side by side to do a direct comparison between the two by keeping all the experimental conditions same. The samples were cleaved *in situ* at ~ 120 K and immediately transferred to STM head, which was precooled at 90 K. Topographic images are acquired in constant current mode with the bias voltage applied to the sample. Differential conductance (dI/dV) spectra were calculated numerically by taking derivative of current-voltage (I-V) measurements. **Figure 4a** gives the STM results of a large-scale image of as-grown parent crystal surface; atomically-resolved images have already been reported [35]. We see hundred nanometers wide flat terraces on both cleaved as-grown and annleaed BaFe$_2$As$_2$ crystalline surfaces; the most common step terrace height is 0.75 nm, which is about half of a unit cell in *c*-axis. Zooming on the flat terrace reveals spatial bright and dark nanoscale regions shown in **Figure 4b**; line profile across the surface indicates a height variation of less than 1 Å (inset). These height variations are too small to be caused by missing atoms in the top layer, and hence most likely related to electronic inhomogeneity due to the fact that STM image is a convolution of spatial variation in the topographic height and the local density of states. The insets of **Figure 4c** show the spatial map of dI/dV spectra at 25 meV, measured in a region with a step edge that appears as a bright line (marked by the arrow) at 90 K. The conductance map on the two flat terraces (as-grown is top; annealed is bottom) looks homogeneous, and an average dI/dV spectra over the whole area exhibits a V-shape (black curves), which is consistent with earlier reported tunneling spectroscopy measurements on the parent compound [36]. However, a detailed analysis reveals different dI/dV spectra in different areas on the surface. Each colored curve is an average of 1250 individual spectra taken in the areas of (200×100) nm marked by corresponding color rectangle in the inset of each figure. For the as-grown crystal (**Figure 4c**, top), in one area, the dI/dV spectra (red curve) exhibits a peak near the Fermi level at 25 mV surrounded by two dips at -125 mV and 85 mV, in contrast to having a single minimum at 65 mV in the adjacent area (blue curve). In another area (green curve), peak at 25 mV is suppressed and result in a more U-shaped dI/dV spectra. The origin of the peak near the Fermi level is unknown at present, and has been also observed in gold-doped BaFe$_2$As$_2$ [37]. Nevertheless, these results indicate that as-grown BaFe$_2$As$_2$ crystal is electronically inhomogeneous at local scales, and the averaging of differential conductance spectra over a large area hides information. A similar analysis done with the dI/dV map taken on the annealed crystal (**Figure 4c**, bottom) shows that annealed sample is comparatively electronically homogeneous as the dI/dV spectra taken at various locations on the surface are qualitatively similar to the one averaged over the whole area (black curve). All the curves show a V-shape with local minimum at 45 mV. However, occasionally a peak type feature appeared near the Fermi level similar to red curve on annealed BaFe$_2$As$_2$ sample with variation in the peak position from -15 mV to 35 mV. In order to observe local chemistry, thin Transmission Electron Microscopy (TEM) specimens of the parent was prepared by focused-ion-beam (FIB), and subsequently by ion milling with liquid nitrogen cooling at a weak beam of 1.5 kV and 3 mA. The conventional TEM and atomic resolution aberration-corrected scanning STEM studies were carried out on an aberration-corrected FEI Titan S 80-300 equipped with a Gatan Image Filter (Quantum-865) at 300 kV. Z-contrast STEM-HAADF (high angle annular dark field) imaging was performed with a probe convergence angle of 30 mrad and an inner collection angle of 65 mrad. STEM images have been reported on various members of iron-based superconductors along different crystallographic projections [38] with high angle annular dark-field detector (HAADF) mode, in which the image's intensity is proportional to $Z^{1.6-2}$. **Figure 4d** shows a typical plane-view of STEM-HAADF image of BaFe$_2$As$_2$. The beam parallel to [001] projection resolves the Ba+As, and Fe columns appearing in bright and medium light, respectively. Although no analyzes of such images were performed along many microns, we suspect that there is some local strain relief for annealed crystalline lattice. For example, our previous study on CaFe$_2$As$_2$ parent annealed (350 °C) crystal with $T_N$=168(1) K showed a strain relief through local 0.2 Å atomic displacements found in-plane HAADF images [39].

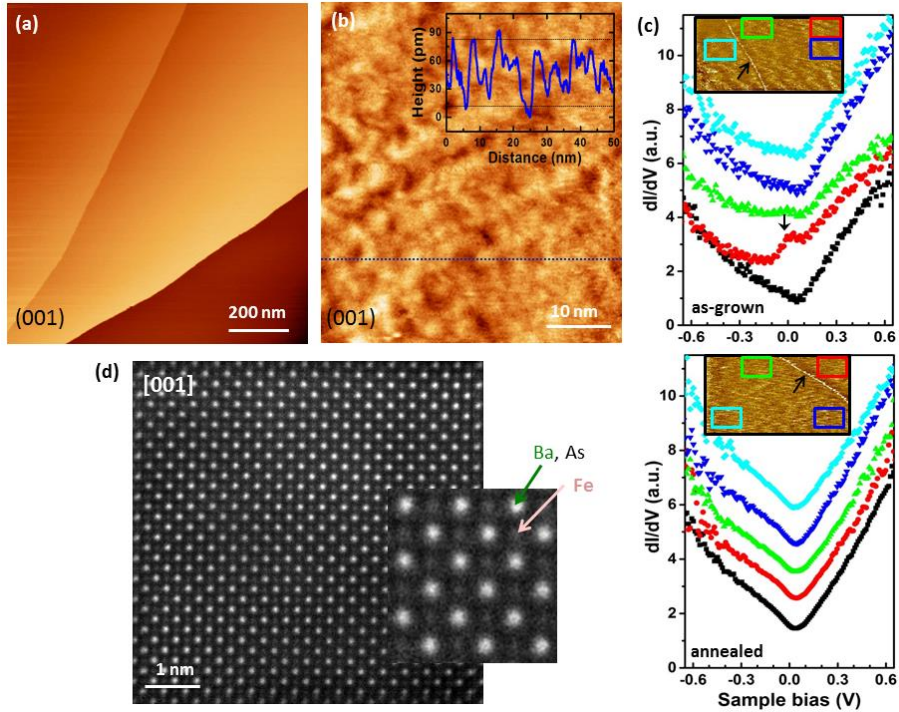

**Figure 4: The improved electronic homogeneity in annealed BaFe$_2$As$_2$ crystals.** For as-grown BaFe$_2$As$_2$, in-plane topographic STM image of (a) cleaved surface showing large atomically flat terraces (V$_s$=-0.5 V, I$_t$=20 pA), (b) a close-up of a flat terrace with tunneling conditions V$_s$=-0.5 V, I$_t$=20 pA, and inset is a line profile taken along the dashed blue line. (c) For as-grown and annealed BaFe$_2$As$_2$, slices of dI/dV maps at 25 meV energy (size for maps is 1000×400 nm, shown in top insets, where spatially averaged differential conductance in smaller locations are marked by corresponding colors and sizes of 200×100 nm); all the dI/dV spectra are offset in y-direction for clarity. In the surface morphology maps, there are step edges that appear as bright lines (marked by arrows). (d) For BaFe$_2$As$_2$, a typical STEM-HAADF image of BaFe$_2$As$_2$ along [001] orientation.

In order to gain insight on $T_N$ changes for BaFe$_2$As$_2$, our first principles calculations here indicate a strong sensitivity of magnetism to structure, consistent with our previously shown strong magnetoelastic coupling [40]. We performed theoretical calculations using the all-electron code WIEN2K [41], within the local density approximation, and using the low-temperature orthorhombic structure. As in [40], we have computed the magnetic ordering energy, defined as the difference in energy ΔE between the ground-state 'stripe' structure and the 'checkerboard' structure of the parent compound and several structural modifications, as presented in **Table 1**. Within a mean-field approximation, this energy difference is proportional to the $T_N$. We note that while the structural modifications described here might be envisioned as application of pressure, one may view them as arising from changes in average lattice parameters caused by synthesis condition or thermal annealing. Indeed, in **Figure 1b** we see an approximate 0.1% decrease in planar lattice parameter (within the tetragonal phase) from annealing. Consistent with this experimental fact, we find that the application of uniform compressive pressure ("hydrostatic"), or along the *c*-axis ("*c*-axis uniaxial") substantially decreases the ordering energy, and hence the Néel temperature in these calculations. For in-plane changes, however, the situation is more complex, and indeed we find that the application of 1% compression along the ferromagnetic Fe-Fe alignment direction ("FM uniaxial"), combined with a 1% tensile stress along the antiferromagnetic Fe-Fe direction, in fact *increases* the ordering energy by 9.2%. Such an increase is roughly consistent with the observed 3% increase in $T_N$ we observe from the annealing, particularly when one considers that the observed decrease in planar lattice parameters would generally be expected to increase the exchange interaction. The suggestion of these results is that slight structural changes can be responsible for the observed $T_N$ change.

**Table 1**: Calculated ordering energies for several structural modifications of $BaFe_2As_2$ parent, as could be produced via hypothetical lattice-parameter changes.

| scenario | $\Delta a$ (%) | $\Delta b$ (%) | $\Delta c$ (%) | $\Delta E$ (meV/Fe) | % change from baseline |
|---|---|---|---|---|---|
| baseline | - | - | - | 49.41 | - |
| hydrostatic | -1 | -1 | -1 | 42.14 | -14.7 |
| $c$-axis uniaxial | +1 | +1 | -2 | 42.29 | -14.6 |
| FM uniaxial | -1 | +1 | - | 53.97 | +9.2 |
| AF uniaxial | +1 | -1 | - | 44.78 | -9.4 |

For superconducting $Ba(Fe_{1-x}Co_x)_2As_2$ crystals annealing produces sharper and higher $T_c$ values, while for optimally-doped crystal critical-current-density ($J_c$) also increases. In the annealed crystal, there is probably nanoscale phase separation of cobalt-depleted and cobalt-rich regions within the same crystal, diminishing the overall distribution of chemical and electronic disorder, leading to the formation of more uniform electronic regions of enhanced or no superconductivity, as is demonstrated below. **Figure 5** shows temperature-dependent results of resistivity ($\rho$) and magnetic susceptibility ($\chi$) for under-doped x=0.023, optimally-doped x=0.063, and over-doped x=0.146 crystals. For x=0.023 (**Figure 5a**), annealing reduces the overall magnitude of $\rho$ and $\chi$, while shifting the $T_N$ value from ~ 90 K to 93 K. For x=0.063 (**Figure 5b**), annealing shows both higher and sharper $T_c$ and a lower normal-state resistivity, and although $\chi$ magnitude is increased slightly at lower temperatures and 1 Tesla, the Meissner fraction is slightly increased at 10 Oe with improvement of $T_c$. For x=0.146 (**Figure 5c**), annealing produces bulk superconductivity at higher temperature of ~ 12 K, even though the absolute values of $\rho$ and $\chi$ increase. A broad superconducting transition temperature in magnetic susceptibility was observed frequently for overdoped crystals, ascribed to most inhomogeneous cobalt-doping composition [19]. For annealed crystal with x=0.063, the rise in $T_c$ is confirmed with anomalies in heat capacity (not shown), although there is no change in the electronic contribution ($\gamma \approx 3$ mJ/K$^2$mol). $J_c$ values were inductively (magnetically) determined by applying the modified critical state model [42,43] to the magnetic hysteresis via the relation $J_c = 20\Delta M/[a(1-a/3b)]$. This relation applies to a rectangular solid with field perpendicular to a face with sides $b > a$. Sample dimensions were $1.52 \times 1.50 \times 0.162$ mm$^3$ for as-grown crystal, and $1.62 \times 1.69 \times 0.345$ mm$^3$ for annealed crystal. Here, $\Delta M = M^- - M^+$ is the magnetic hysteresis, where $M^-(M^+)$ is the magnetization at temperatures $T$ measured in decreasing (increasing) field $H$ history. With the superconducting crystals in perpendicular field geometry, the flux density $B$ can be replaced by $\mu_o H$ to a close approximation. Fields in the range 0 - 6.5 T and in the $c$-direction were applied at different fixed temperatures (5 to 18.5 K) and the moment generated by the induced flowing current in the crystal was measured; before beginning measurement, the magnet was reset to eliminate any trapped flux and assure zero $H$. **Figure 6a** shows improved $J_c$ values for annealed crystals. $J_c$ values are calculated at two different temperatures below $T_c$, and as a function of applied field $H$. The 'fishtail' $J_c$ peak effects have been observed and reported in similar crystals [44-47]; $J_c$ is fairly constant at very low fields (up to 0.07 T), then starts to drop off (from about 0.08 to 0.6 T), and can gradually increase and fall again. This feature may indicate the presence of nanoscale phase separation into regions of weaker superconductivity, perhaps caused by an inhomogeneous distribution of the cobalt [48]. Also for both crystals, $J_c$ does not fall off too rapidly with the application of magnetic field; this implies its weaker dependence on $H$. The $J_c$ in the annealed crystal is more than four times larger than that measured in the as-grown crystal at very low fields, while it is about three times larger at intermediate and higher fields. The annealed sample has a self-field $J_c$ of about 1 MA/cm$^2$ at 5 K, which compares favorably with those measured in Co-doped $BaFe_2As_2$ epitaxial films deposited on LSAT and MgO substrates at 1-4 MA/cm$^2$ at 4 K [49]. Other $J_c$ values of 0.4 MA/cm$^2$ at 4.2 K [44] and 0.26 MA/cm$^2$ at 5 K [46] have also been reported. Further insight into the pinning strength can be achieved when the temperature dependence of $J_c$ is analyzed, shown at an applied field $H = 0.2$ T (**Figure 6b**). Superconductors with a weak pinning behavior such as ours usually exhibit an exponential decrease in $J_c$ with temperature at low applied fields, owing to the low effectiveness of point-like defects against thermal activation of vortices [50]. Strong pinning mechanisms, such as correlated disorder pinning instead exhibit more smooth temperature dependence. The enhancement in $J_c$ in the annealed sample is attributable to cobalt clustering seen as defects in the plane view (seen in microscopy images, below), giving rise to better pinning.

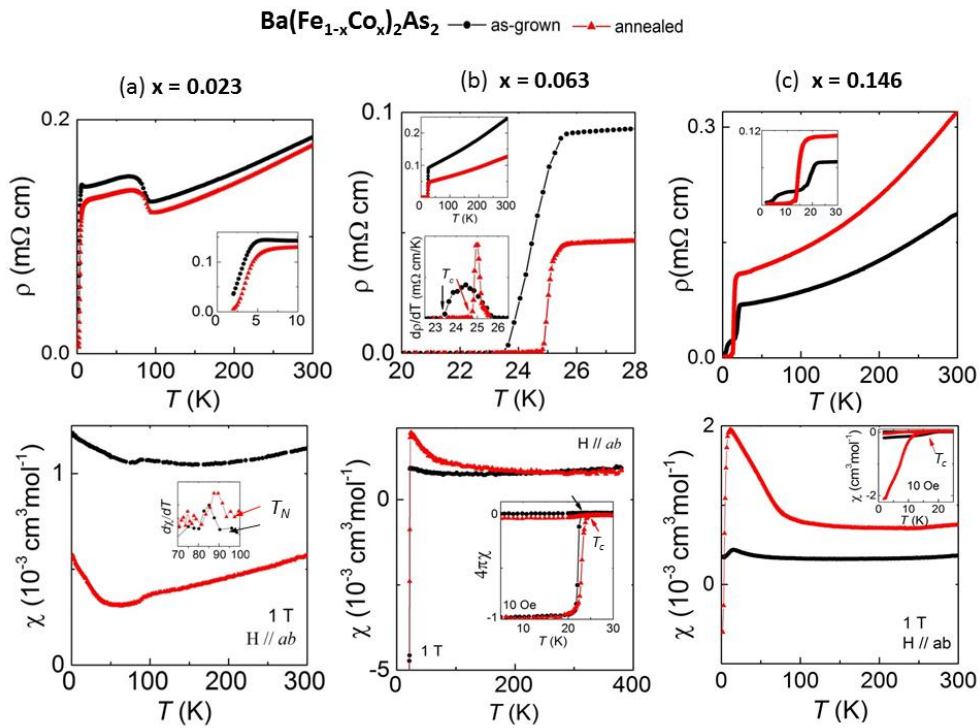

**Figure 5: The improvement in superconducting critical temperature ($T_c$) signals upon annealing Ba(Fe$_{1-x}$Co$_x$)$_2$As$_2$ crystals.** For Ba(Fe$_{1-x}$Co$_x$)$_2$As$_2$ crystals of as-grown (in black) and annealed (in red), temperature dependence of resistivity (ρ; top row) and magnetic susceptibility (χ; bottom row) for (a) x = 0.023, (b) x = 0.063, and (c) x = 0.146.

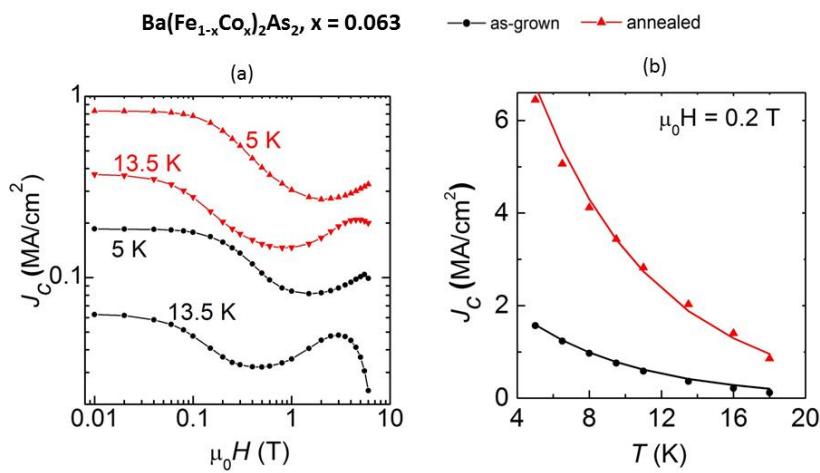

**Figure 6: The increase in superconducting critical current density ($J_c$) for annealed optimally-doped Ba(Fe$_{1-x}$Co$_x$)$_2$As$_2$ crystal.** For x = 0.063 and as-grown (in black) and annealed (in red) crystals, (a) field dependence of critical current density ($J_c$) below $T_c$, and (b) temperature dependence of $J_c$ at 0.2 Tesla.

**Figure 7** gives evidence of chemical clustering in annealed superconducting crystal. Thin TEM specimens of Ba(Fe$_{1-x}$Co$_x$)$_2$As$_2$ with x=0.063 were also prepared as described above. Also, electron-energy-loss-spectroscopy (EELS) data were collected in STEM mode using a dispersion of 0.25eV per channel, a 5 mm aperture, and a collection angle of 40 mrad. Some EEL spectra and EELS mapping were acquired on a Nion UltraSTEM 100TM operating at 100 kV [51] equipped with a 5th order probe aberration corrector and a Gatan Enfina EEL spectrometer. STEM images of several micron areas on two pieces of each of as-grown versus annealed x= 0.063 were comparable. STEM results can give bulk chemical nature, as they are the average signal of many atomic unit cells along the beam direction. As shown in **Figure 7a**, the low-magnification image along [001] that which STM/S were analyzed, show differences: uniform contrast in as-grown crystal indicates chemical homogeneity (top), however, the nano-clusters (dark regions) in size of around 10 nm or smaller in annealed crystal indicates chemical non-uniformity (bottom) that can be separated by more than 100 nm. Typical atomic resolution image on either crystal is shown in the inset of top image, and shows no visible defects. EELS results are shown in **Figure 7b**, indicating chemical inhomogeneity for annealed crystal and two types of nano-clusters: one type of nano-cluster regions shows weaker Ba signal (top spectra; collected on Titan at 300 kV), while the other one reveals stronger Ba signal (bottom spectra; collected on Nion 100 at 100 kV). The EELS mapping of the latter type was performed and is shown in **Figure 7c**, demonstrating such chemical inhomogeneity. Since Co $L_{2,3}$ edge overlaps with Ba $M$-edge, the weaker or stronger Ba $M$-edge in the nano-clusters could be the result of less or more cobalt. In support of this and as evident in refinements of unit cells above, $c$-parameter should increase if there is less cobalt substituted in the 122 structure, i.e. it is possible that cobalt precipitates out by forming grain boundaries, leaving the main matrix of annealed crystal having less cobalt.

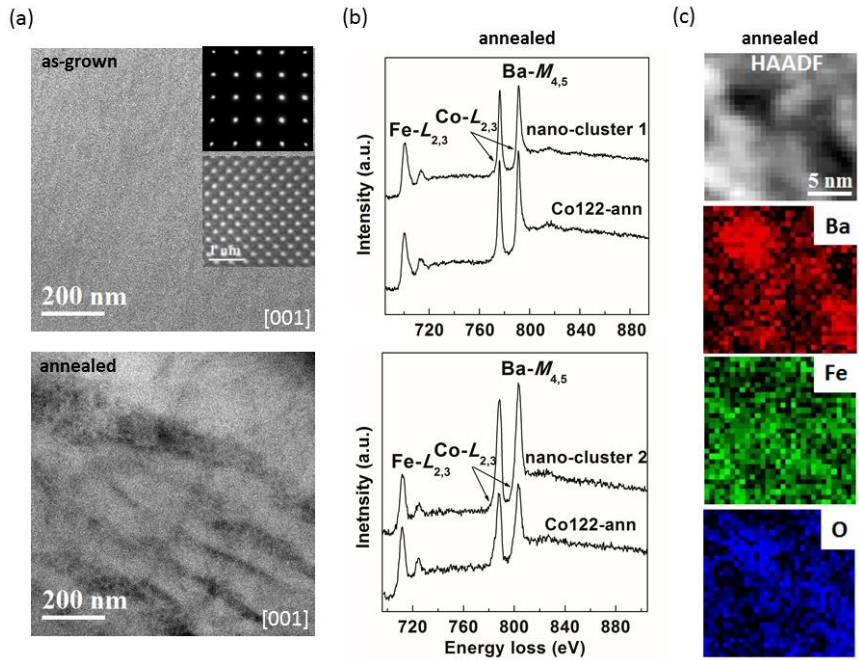

**Figure 7: Evidence of chemical clustering in annealed optimally-doped Ba(Fe$_{1-x}$Co$_x$)$_2$As$_2$ crystal:** (a) Typical STEM HAADF images along [001] showing chemical homogeneity for as-grown (top) versus nano-clusters in annealed (bottom) crystals, the insets are typical electron diffraction patterns and atomic resolution image. For annealed crystals, two different types of nano-clusters showing chemical inhomogeneity as revealed by (b) EEL spectra and (c) EELS mapping for one type of nano-clusters.

Such chemical non-uniformity should effect local electronic structures. Although the surfaces of BaFe$_2$As$_2$ are known to reconstruct, the superconducting gaps from different surface reconstruction are expected to be almost identical since the superconductivity is a global property and ~3 nm coherence lengths are found for these materials [52-54]. In fact, for Ba$_{0.5}$Sr$_{0.5}$(Fe$_{1-x}$Co$_x$)$_2$As$_2$ with x=0.073 as-grown crystal ($T_c$=17K), scanning tunneling microscopy/spectroscopy (STM/S) results of at 4.3 K found inhomogeneous gap values from about 3 meV down to 0 meV [20]. For Ba(Fe$_{1-x}$Co$_x$)$_2$As$_2$ crystals, **Figure 8** demonstrates STM/S results at 4.2 K and the smaller variability in superconducting gap maps for the annealed x=0.063 crystal. For x=0.063 crystals, both as-grown and annealed pieces were cleaved *in situ* at ~ 100 K and then immediately transferred to STM head which was precooled at 4.2 K. Topographic images were acquired in constant current mode with the bias voltage applied to the sample. The dI/dV spectroscopies were obtained using the lock-in technique with bias modulation $V_{rms}$ = 0.5 mV at 973 Hz. **Figure 8a** shows the topographic images acquired from the set of crystals. While the large-scale morphologies of the surfaces are essentially the same for the two, the detailed atomic-level structures may show some discrepancies from cleavages, consistent with the reported cleavage dependent controversies from the literature [52]. To reveal the real space superconductor gap information from the crystals, the current-imaging-tunneling spectroscopy (CITS) were carried out; **Figure 8b** gives the normalized dI/dV spectroscopy averaged over the whole areas of (a). By fitting the dI/dV spectroscopies using the standard functional form from Dynes *et al* [55] the gap Δ maps can be deducted from the CITS images, as shown in **Figure 8c**. In the map of the as-grown crystal, much more dispersed electronic regions with different gaps are seen, while in the map of the annealed crystal, there are larger areas of non-superconducting clusters and the green regions with gap ~5 meV is more prominent and are connected. The statistic histogram distributions of the superconducting gaps in **Figure 8d** show that the peaks of the gap are located at 6.6 meV and 5.0 meV, respectively for as-grown versus annealed, with full-width-at-half-maximum (FWHM) of 4.9 and 4.0. The main difference between the two gap distributions is a much higher probability at Δ=0 for the annealed crystal, supporting the higher pinning sites suggested by higher $J_c$. The smaller FWHM for the annealed sample means narrower gap distribution, consistent with the sharper superconductivity transition, shown in bulk properties. The extracted corresponding gap-to-$T_c$ ratios $R=2\Delta/k_bT_c$ from the peaks of the gaps are 6.7 and 4.6, respectively, using $T_c$=23 K and 25 K for the as-grown and annealed crystals. Although in the weak coupling *s*-wave BCS theory the *R* is a constant of 3.53, the *d*-wave symmetry of the order parameter in cuprate superconductors makes the ratio to a larger value of 4.28 and higher [56]. Furthermore, recent studies found *R* of electron or hole doped $BaFe_2As_2$ can vary from 2.2 to 10.3 [53,57-61]. Therefore, in the annealed crystal, there is evidence of nanoscale phase separation (cobalt clustering) into cobalt rich and deleted regions reducing the local lattice strain introduced by doping directly within the Fe planes, giving sharper and slightly higher $T_c$. Such nano-size chemical phase separation of cobalt may be responsible for the higher $J_c$ value.

With regards to the effects of annealing on the superconducting state, it is rather remarkable that annealing increases $T_c$ while significantly decreasing the average superconducting gap size. Usually a smaller gap correlates with a decreased $T_c$, given that most theories predict a constant ratio of the gap to $T_c$, within a weak-coupling regime. One clue to the origin of this unusual behavior can be found in the averaged dI/dV curves. One notes that the coherence peaks located at approximately ±5 mV are significantly sharper in the annealed sample, and reach greater heights and form better-connected percolating regions. In these samples, these features are ultimately controlled by two factors: the amount of quasiparticle scattering in the sample (discussed below), as well as by the distribution of gap values. A sample with a narrower distribution of gap values, as we observe in the annealed sample, will also tend to have higher and sharper coherence peaks when the resulting dI/dV curves are averaged over many locations. Quasiparticle scattering is typically modeled by the parameter Γ introduced by Dynes. In a pure *s*-wave sample, the scattering modeled by Γ does not typically affect $T_c$ substantially, if the scattering originates in a non-magnetic manner (Anderson's theorem). However, for more complex pairing symmetries such as $s_{+/-}$ or *d*-wave, such scattering can greatly impact $T_c$, with the magnitude of the effect dependent both upon the specific pairing symmetry as well as the strength of the scattering. We therefore suggest, in addition to inducing a narrower distribution of gap values, that the annealing in some manner reduces the quasiparticle scattering, and thereby the pair-breaking effects of such scattering, thus raising $T_c$. Complicating such an interpretation, however, is the substantially larger fraction (12%) of small (<1 meV) gap regions in the annealed state, relative to the 5% in the as-grown crystal. One would typically associate a smaller Γ with a more homogeneous chemical and electronic structure, yet this larger small-gap fraction argues in the opposite direction. It is possible, though clearly unproven here, that these small gap regions somehow play an important role in the global superconducting behavior, such as by donating charge to the system, despite apparently impeding superconductivity at the local level. In fact, the enhancement of superconductivity at the boundary between strongly underdoped and overdoped regions has been observed in the past. For example in the cuprates, a record $T_c$ of 50 K was reported in bilayer of strongly overdoped non-superconducting $La_{1+x}Sr_xCuO_4$ (x=0.45), and the top layer of underdoped insulating $La_2CuO_4$. The enhanced superconductivity was confined to a very thin (~ 2 unit cells) interfacial layer, and believed to be the transfer of charge from the overdoped to the underdoped layer across the interface, providing optimal doping without introducing chemical disorder. Similar to this case, cobalt clustering in the annealed crystal may create regions made up of cobalt-enriched and cobalt-depleted regions with reduced quenched disorder analogous to the case of under/overdoped $La_2CuO_4$ sublattices [62].

This research has allowed us to gain understanding of how competing and different distributions of local disorder can cluster in a way to control antiferromagnetic ordering temperature or cause percolative bulk zero resistance below a temperature in single crystals. In the present work, we have looked at a few crystal compositions within $Ba(Fe_{1-x}Co_x)_2As_2$ system, and refined the bulk compositions and structures using X-ray spectroscopy and diffraction techniques, and also used microscopy and spectroscopy experiments to investigate local chemical and electronic structure disorder within crystals. Higher $T_N$ seems to be arising from more globally ordered lattice with shorter average *a*-lattice parameter and homogeneous electronic structure upon annealing, and higher $T_c$ and $J_c$ in optimally-doped crystal is correlated to less electronic structure variation, nanoscale phase separation and larger average *a*-lattice parameter. Hence, annealing improves transition temperatures in $Ba(Fe_{1-x}Co_x)_2As_2$ crystals by removing the local-lattice strain and much of the electronic-structure disorder/disparity; this is simply summarized in **Figure 9**.

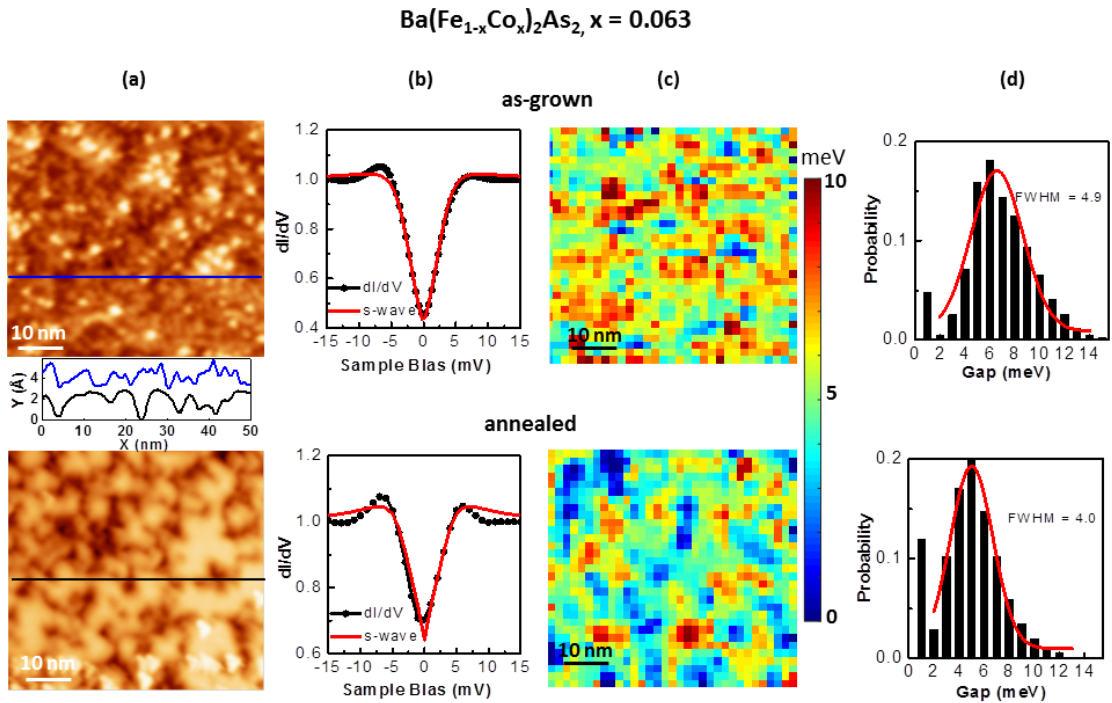

**Figure 8: Narrower superconducting gap distributions, and more uniform percolative superconducting electronic clusters for annealed optimally-doped $Ba(Fe_{1-x}Co_x)_2As_2$ crystal.** For crystal with x = 0.063, and as-grown (top row) and annealed (bottom row) crystals cleaved in-situ at 100 K: (a) topographic STM images with sample-bias voltage $V_b$= 20 mV and tunneling current $I_t$= 100 pA for as-grown crystal, and $V_b$= 25 mV and $I_t$= 100 pA for annealed crystal, and 50 nm × 50 nm areas. (b) normalized dI/dV spectroscopy averaged over the whole areas of (a) (1024 datasets), with s-wave fitting of the data in red. (c) Superconducting gap maps derived from dI/dV spectra from (a) with modulation $V_{rms}$ = 0.5 mV at 973 Hz. (d) Histogram distributions of the gap Δ of (c) respective gap maps, with Gaussian fitting in red. Line profiles along the blue and black lines of (a) are shown in-between the two images. The two lines are offset for clarity.

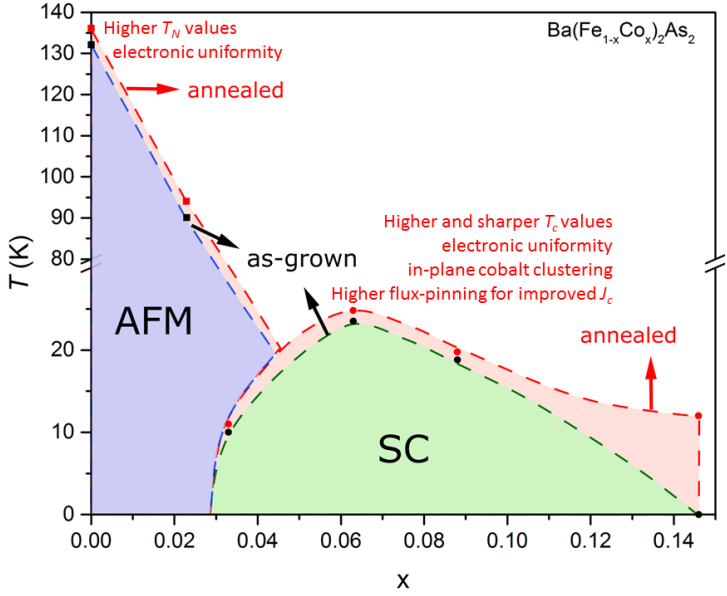

**Figure 9: Annealing improves quantum transition temperatures in Ba(Fe$_{1-x}$Co$_x$)$_2$As$_2$ crystals.** For all crystals, the overall lattice strain can be removed while the electronic structures become more uniform. While the in-plane cobalt chemical clustering can improve pinning, it seems to also raise the overall superconducting (SC) dome by giving paths for better percolation. Homogeneity improves antiferromagnetic transition temperatures (AFM).

**Acknowledgements**

The research is primarily supported by the U.S. Department of Energy (DOE), Office of Science, Basic Energy Sciences (BES), Materials Science and Engineering Division. The spectroscopy and microscopy work was performed through user projects supported by the ORNL's Center for Nanophase Materials Sciences (CNMS), which is sponsored by an Office of Science User Facility, DOE. A.O.I. work was supported by the U.S. Department of Energy, Office of Science, Office of Workforce Development for Teachers and Scientists (WDTS) under the VFP. The single-crystal neutron and X-ray diffraction work at ORNL were sponsored by the Scientific User Facilities Division, Office of BES, DOE. Z.W. work was supported by the National Natural Science Foundations of China (no. 61227009). The authors acknowledge assistance by J. Mitchell, Z. Sims, and M. Susner for some of the sample preparations, and also L. Walker for composition line analysis on a crystal. We also thank C. Hoffman for discussions.

**References**

[1] A.S. Sefat and D.J. Singh, *MRS Bull.* **2011**, *36*, 614.
[2] M.B. Maple, *Phys. B Condens. Matter* **1995**, *215*, 110.
[3] J. Orenstein and A.J. Millis, *Science* **2000**, *288*, 468.
[4] K. Ishida, Y. Nakai, and H. Hosono, *J. Phys. Soc. Jpn.* **2009**, *78*, 62001.
[5] H. Alloul, J. Bobroff, M. Gabay, and P.J. Hirschfeld, *Rev. Mod. Phys.* **2009**, *81*, 45.
[6] A.S. Sefat, *Rep. Prog. Phys.* **2011**, *74*, 124502.
[7] P. Dai, J. Hu, and E. Dagotto, *Nat. Phys.* **2012**, *8*, 709.
[8] L.M.N. Konzen, A.S. Sefat, "Lattice parameters guide superconductivity in iron-arsenides," *J. Phys.: Condens. Matter* 2017 (accepted).
[9] Lumsden, M.D., Christianson, A.D., *J. Phys.: Condensed Matter* **2010**, *22*, 203203.
[10] A.S. Sefat, R. Jin, M.A. McGuire, B.C. Sales, D.J. Singh, D. Mandrus, *Phys. Rev. Lett.* **2008**, *101*, 117004.
[11] A.S. Sefat, M.A. McGuire, R. Jin, B.C. Sales, D. Mandrus, F. Ronning, E.D. Bauer, Y. Mozharivskyj, *Phys. Rev. B* **2009**, *79*, 094508.

[12] C.R. Rotundu, B. Freelon, T.R. Forrest, S.D. Wilson, P.N. Valdivia, G. Pinuellas, A. Kim, J.W. Kim, Z. Islam, E. Bourret-Courchesne, N.E. Phillips, R.J. Birgeneau, *Phys. Rev. B* **2010**, *82*, 144525.
[13] B. Saparov, A.S. Sefat, *Dalton Trans.* **2014**, *43*, 14971.
[14] M. Rotter, M. Tegel, D. Johrendt, I. Schellenberg, W. Hermes, R. Poettgen, *Phys. Rev. B* **2008**, *78*, 020503.
[15] J. Gillett, D. Sitikantha, P. Syers, A.K.T. Ming, J.I. Espeso, C.M. Petrone, S.E. Sebastian, S.E., arXiv:1005.1330, **2010**.
[16] K. Gofryk, B. Saparov, T. Durakiewicz, A. Chikina, Danzenbacher, D.V. Vyalikh, M.J. Graf, A.S. Sefat, *Phys. Rev. Lett.* **2014**, *112*, 186401.
[17] K. Gofryk, A.B. Vorontsov, I. Vekhter, A.S. Sefat, T. Imai, E.D. Bauer, J.D. Thompson, F. Ronning, *Phys. Rev. B* **2011**, *83*, 064513.
[18] K. Gofryk, A.S. Sefat, M.A. McGuire, B.C. Sales, D. Mandrus, T. Imai, J.D. Thompson, E.D. Bauer, F. Ronning, *J. Phys.* **2011**, *73*, 012094.
[19] D.L. Sun, J.Z. Xiao, C.T. Lin, *Journal of Crystal Growth* **2011**, *321*, 55.
[20] J.E. Mitchell, B. Saparov, W. Lin, S. Calder, Q. Li, S.V. Kalinin, M. Pan, A.D. Christianson, A.S. Sefat, *Phys. Rev. B* **2012**, *86*, 174511.
[21] T. Berlijn, C.H. Lin, W. Garber, W. Ku, *Phys. Rev. Lett.* **2012**, *108*, 207003.
[22] A.S. Sefat, *Rep. Prog. Phys.* **2011**, *74*, 124502.
[23] C. Dhital, Z. Yamani, W. Tian, J. Zeretsky, A.S. Sefat, Z. Wang, R.J. Birgeneau, S.D. Wilson, *Phys. Rev. Lett.* **2012**, *108*, 087001.
[24] K. Gofryk, M. Pan, C. Cantoni, B. Saparov, J.E. Mitchell, A.S. Sefat, *Phys. Rev. Lett.* **2014**, *112*, 047005.
[25] L. Akselrud and Y. Grin, *J. Appl. Crystallogr.* **2014**, *47*, 803.
[26] N. Ni, M.E. Tillman, J.Q. Yan, A. Kracher, S. T. Hannahs, S.L. Bud'ko, P.C. Canfield, *Phys. Rev. B* **78** (2008), 214515.
[27] *Rigaku/MSC*, **2014**. *CrystalClear 2.0. Rigaku/MSC Inc., The Woodlands, Texas, USA.* (n.d.).
[28] G.M. Sheldrick, *Acta Crystallogr. Sect. C Struct. Chem.* **2015**, *71*, 3.
[29] E.D. Mun, S.L. Bud'ko, N. Ni, A.N. Thaler, P.C. Canfield, *Phys. Rev. B* **2009**, *80*, 054517.
[30] F. Rullier-Albenque, D. Colson, A. Forget, H. Alloul, *Phys. Rev. Lett.* **2009**, *103*, 057001.
[31] B.C. Chakoumakos, H. Cao, F. Ye, A.D. Stoica, M. Popovici, M. Sundaram, W. Zhou, J.S. Hicks, G.W. Lynn, and R.A. Riedel, *J. Appl. Crystallogr.* **2011**, *44*, 655.
[32] K. Marty, A.D. Christianson, C.H. Wang, M. Matsuda, H. Cao, L.H. VanBebber, J.L. Zarestky, D.J. Singh, A.S. Sefat, M.D. Lumsden, *Phys. Rev. B* **2011**, *83*, 060509.
[33] L. Li, H. Cao, M.A. McGuire, J.S. Kim, G.R. Stewart, A.S. Sefat, *Phys. Rev. B* **2015**, *92*, 094504.
[34] T.R. Forrest, P.N. Valdivia, C.R. Rotundu, E. Bourret-Courchesne, R.J. Birgeneau, *J. Phys.: Condens. Matter* 28 (2016), 115702.
[35] R. Jin, M.H. Pan, X.B. He, G. Li, D. Li, R.W. Peng, J. R. Thompson, B.C. Sales, A.S. Sefat, M.A. McGuire, D. Mandrus, J.F. Wendelken, V. Keppens, E.W. Plummer, *Supercon. Sci. Tech.* **2010**, *23*, 054005.
[36] F. Massee, Y. Huang, R. Huisman, S. de Jong, J.B. Goedkoop, M.S. Golden, *Phys. Rev. B* **2009**, *79*, 220517.
[37] M. Ziatdinov, A. Maksov, L. Li, A.S. Sefat, P. Maksymovych, S.V. Kalinin, *Nanotechnology* **2016**, *27*, 475706.
[38] C. Cantoni, J.E. Mitchell, A.F. May, M.A. McGuire, J.C. Idrobo, T. Berlijn, E. Dagotto, M.F. Chisholm, W. Zhou, S.J. Pennycook, A.S. Sefat, B.C. Sales, *Adv. Mater.* **2014**, *26*, 6193.
[39] B. Saparov, C. Cantoni, M. Pan, T.C. Hogan, W. Ratcliff, S.D. Wilson, K. Fritsch, B.D. Gaulin, A.S. Sefat, *Sci. Rep.* **2014**, *4*, 4120.
[40] A.S. Sefat, L. Li, H.B. Cao, M.A. McGuire, B. Sales, R. Custelcean, D.S. Parker, *Sci. Rep.* **2016**, *6*, 21660.
[41] Blaha, P. *et al.* WIEN2k, An augmented plane wave + local orbitals program for calculating crystal *properties* (Karlheinz Schwarz, Techn. Universitat Wien, Austria, **2001**).

[42] C.P. Bean, *Rev. Mod. Phys.* **1964**, *36*, 31.
[43] C. P. Bean, *Phys. Rev. Lett.* **1962,** *8*, 250.
[44] A. Yamamoto, J. Jaroszynski, C. Tarantini, L. Balicas, J. Jiang, A. Gurevich, D.C. Larbalestier, R. Jin, A.S. Sefat, M.A. McGuire, B.C. Sales, D.K. Christen, D. Mandrus, *Appl. Phys. Lett.* **2009**, *94*, 62511.
[45] H. Yang, H. Luo, Z. Wang, and H.-H. Wen, *Appl. Phys. Lett.* **2008**, *93*, 142506.
[46] R. Prozorov, N. Ni, M.A. Tanatar, V.G. Kogan, R.T. Gordon, C. Martin, E.C. Blomberg, P. Prommapan, J.Q. Yan, S.L. Bud'ko, and P.C. Canfield, *Phys. Rev. B* **2008**, *78*, 224506.
[47] D.L. Sun, Y. Liu, and C.T. Lin, *Phys. Rev. B* **2009**, *80*, 144515.
[48] M. Putti, I. Pallecchi, E. Bellingeri, M.R. Cimberle, M. Tropeano, C. Ferdeghini, A. Palenzona, C. Tarantini, A Yamamoto, J. Jiang, J. Jaroszynski, F. Kametani, D. Abraimov, A. Polyanskii, J.D. Weiss, E.E. Hellstrom, A Gurevich, D.C. Larbalestier, R. Jin, B.C. Sales, A.S. Sefat, M.A. McGuire, D. Mandrus, P. Cheng, Y. Jia, H.H. Wen, S Lee, and C.B. Eom, *Supercond. Sci. Technol.* **2010**, *23*, 34003.
[49] T. Katase, H. Hiramatsu, T. Kamiya, and H. Hosono, *Appl. Phys. Express* **2010**, *3*, 63101.
[50] G. Blatter, M.V. Feigel'man, V.B. Geshkenbein, A.I. Larkin, and V.M. Vinokur, *Rev. Mod. Phys.* **1994**, *66*, 1125.
[51] O.L. Krivanek, G.J. Corbin, N. Dellby, B.F. Elston, R.J. Keyse, M.F. Murfitt, C.S. Own, Z.S. Szilagyi, and J.W. Woodruff, *Ultramicroscopy* **2008**, *108*, 179.
[52] J.E. Hoffman, *Rep. Prog. Phys.* **2011**, *74*, 124513.
[53] F. Massee, Y.K. Huang, J. Kaas, E. van Heumen, S. de Jong, R. Huisman, H. Luigjes, J.B. Goedkoop, and M.S. Golden, *Europhys. Lett.* **2010**, *92*, 57012.
[54] Yin, Y.; Zech, M.; Williams, T. L.; Wang, X. F.; Wu, G.; Chen, X. H.; Hoffman, J. E., *Phys. Rev. Lett.* **2009**, *102*, 097002.
[55] R.C. Dynes, V. Narayanamurti, and J.P. Garno, *Phys. Rev. Lett.* **1978**, *41*, 1509.
[56] C. Pakokthom, B. Krunavakarn, P. Udomsamuthirun, and S. Yoksan, *J. Supercond.* **1998**, 11, 429.
[57] F. Massee, Y. Huang, R. Huisman, S. de Jong, J.B. Goedkoop, and M.S. Golden, *Phys. Rev. B* **2009**, *79*, 220517.
[58] L. Shan, Y.-L. Wang, J. Gong, B. Shen, Y. Huang, H. Yang, C. Ren, and H.-H. Wen, *Phys. Rev. B* **2011**, *83*, 60510.
[59] M.L. Teague, G.K. Drayna, G.P. Lockhart, P. Cheng, B. Shen, H.-H. Wen, and N.-C. Yeh, *Phys. Rev. Lett.* **2011**, *106*, 87004.
[60] Y. Yin, M. Zech, T.L. Williams, X.F. Wang, G. Wu, X.H. Chen, and J.E. Hoffman, *Phys. Rev. Lett.* **2009**, *102*, 97002.
[61] H. Zhang, J. Dai, Y. Zhang, D. Qu, H. Ji, G. Wu, X.F. Wang, X.H. Chen, B. Wang, C. Zeng, J. Yang, and J.G. Hou, *Phys. Rev. B* **2010**, *81*, 104520.
[62] Gozar A., Logvenor G., Fitting Kourkoutis L., Bollinger A.T., Giannuzzi L.A., Muller D.A., Bozovic I. High-temperature interface superconductivity between metallic and insulating copper oxides. *Nature* **2008**, *455*, 782.